\documentstyle{article}

\title{Semiclassical Gravity and Large--Scale Structure}
\author{C. Bertoni \and E. Carretti \and F. Finelli
\and A. Messina \and  G. Venturi}
\date{}

\begin{document}

\maketitle
\begin{center}
{\small\it
Dipartimento di Fisica, Universit\`a degli Studi di Bologna} \\
{\small\it
Istituto Nazionale di Fisica Nucleare, Sezione di Bologna} \\
{\small\it
via Irnerio, 46} \\
{\small\it
40126 Bologna -- Italy}
\end{center}

\bigskip

\begin{abstract}
The possible cosmological effects of primordial fluctuation
corrections to the evolution
equation of matter obtained from the Wheeler--De Witt equation are
explored. In particular, both
the metric and a scalar matter field are expanded around their
homogeneous values and the corrections induced on the scalar field
fluctuation spectrum are perturbatively
estimated. Finally, results of a preliminary numerical simulation to
investigate the effects on large--scale structure formation are presented.
\end{abstract}

The matter--gravity system may be studied quantum--mechanically through
the canonical quantization of gravity
within the superspace approach \cite{wheeler}
\cite{dewitt}. One then obtains the Wheeler--De Witt (WD) equation and a
matter--gravity wave
function which, in analogy with the Born--Oppenheimer (BO) studies of molecules
\cite{messiah}, may be factorized into two parts: one involving only
gravitational degrees of freedom and the other
involving both the gravitational and the matter
degrees of freedom. Correspondingly, the WD equation
in which time is absent is split into two pieces: one describing
gravitation in an effective potential given by
the mean energy--momentum tensor of matter and the other describing
the matter whose evolution is parametrized by time which is
derived from the semiclassical approximation to the gravitational
wave function \cite{banks} \cite{brout}.
The above is contingent on the hypothesis that
the Planck mass is much larger than the mass of any matter field or
any inverse length scale used to describe matter.

The above approach has been examined in detail within the context
of a minisuperspace model with matter \cite{broutventuri}
with the view of understanding
under what conditions the BO approximation is valid and quantum
cosmology leads to the usual physics (Schr\"{o}dinger equation for
matter and
Einstein classical equations for gravity on scales larger than the
Planckian). It was found that such was the case in an inflationary
scenario \cite{broutenglert}
and after ten or more Planck times the usual physics ensues.
Such a result was obtained by estimating perturbatively
the fluctuation corrections
due to quantum gravitational effects both on the equation of motion
for the purely gravitational part and that for matter,
since the actual solution of the coupled nonlinear equations is
extremely complicated.

The fluctuation corrections considered were a consequence of matter--gravity
forming a closed system: fluctuations of gravitational origin generate
corresponding fluctuations in the evolution of matter and it is
such fluctuations that are neglected in the BO approximation.
Henceforth we shall refer to them as fluctuation corrections.

It has been previously suggested that inflation could in principle
provide a causal mechanism for the origin of density perturbations
that later grow to form large--scale structures \cite{guthpi}, thus
we feel it is of interest to explore
the influence of the above fluctuation corrections in such a context.
In order to do so we generalize the previous analysis \cite{broutventuri}
to include
perturbations both for the three--metric and for the scalar field
associated with inflation. In particular, both the perturbations
of the
three metric and the scalar field (which represents matter and
whose non--zero vacuum expectation value has been incorporated in
the cosmological constant) are expanded in terms of scalar harmonics
on the three
sphere and just retained to second order \cite{hallhaw}.
We then consider the resulting coupled equations in the presence of a positive
cosmological constant (de Sitter)
with the scope of obtaining information on the large--scale
structures.

Since we can not actually solve the nonlinear coupled matter--gravity
system we do not know the effect of fluctuation corrections on
the scalar field energy and the resulting {\em distorted}
scalar field spectrum.
However we can estimate the fluctuation corrections perturbatively using, as
unperturbed solution, the matter wave function obtained on neglecting them
(BO approximation). One then assumes that the distorted scalar
field spectrum is reflected by the relative weight of the modes (unperturbed
plus contributions due to fluctuation corrections) at a given instant, say
when each mode exits the horizon during the inflationary era \cite{wada}.
It is further clear that the relative strength
of the diverse corrections besides depending
on the moment at which they are {\em frozen} will also depend on the initial
value of the cosmological constant, in any case the fluctuation
corrections must always be less than the unperturbed result otherwise
our perturbative approach is meaningless.

In order to illustrate our approach we recall the perturbed minisuperspace
results previously obtained \cite{hallhaw}. The spatial metric
$h_{\alpha \beta}$ is given by

\begin{equation}
 h_{\alpha \beta}=a^2(\eta) \left( \Omega_{\alpha \beta}
+ \epsilon_{\alpha \beta} \right) \,,
\end{equation}

\noindent
where $a$ is the Robertson--Walker scale factor, $\eta$ the
conformal time, $\Omega_{\alpha \beta}$
the metric on a sphere of unit radius and
$\epsilon_{\alpha \beta}$ a perturbation to it which can be
expanded in terms of the spherical harmonics of $S^3$ \cite{hallhaw}:

\begin{eqnarray}
\label{vac}
\epsilon_{\alpha\beta}=\sum_{n,l,m} \left[ 6^{1 \over 2} a_{nlm} {1 \over 3}
\Omega_{\alpha\beta}Q^n_{lm} + 6^{1\over 2} b_{nlm}(P_{\alpha\beta})^n_{lm}
 \right] \,,
\end{eqnarray}

\noindent
where the coefficients $a_{nlm}$, $b_{nlm}$ are functions of $\eta$ but
not of the spatial coordinates. Similarly, one may also expand the lapse ($N$),
shift ($N_\alpha$) functions and the scalar
(matter) field $\Phi$:

\begin{equation}
N= a \left[ 1+6^{- {1\over 2}} \sum_{n,l,m} g_{nlm}
Q_{lm}^n \right] \,,
\end{equation}

\begin{equation}
N_{\alpha}=a \sum_{n,l,m}  6^{-{1\over 2}} k_{nlm} (P_{\alpha})^n_{lm} \,,
\label{eq:vettsp}
\end{equation}

\begin{equation}
\label{vac2}
\Phi = \phi (\eta) + \sum_{n,l,m} f_{nlm} Q^n_{lm} \,.
\end{equation}

\noindent
and we have just considered terms in the expansions eqs.
(\ref{vac})--(\ref{vac2})
associated with scalar harmonics $Q^{n}_{lm}$ on the three sphere
(or their covariant derivatives \cite{harmo}) since it is
these contributions which are relevant for the non--homogeneous part of the
scalar
field in the semiclassical limit \cite{hallhaw}. Henceforth, for the sake
of brevity, we shall only exhibit the
index $n$, other indices ($l$, $m$) being understood.

One may now substitute the above into the total Hamiltonian
density (gravitation plus matter) and perform the spatial integration
keeping the minisuperspace variable $a$ to all orders
and the perturbations $\phi$, $a_n$, $b_n$, $f_n$, $k_n$
only to second order. These now become the dynamical variables and
one may now quantize canonically obtaining the following WD equation
and momentum conditions:

\begin{equation}
\left( \hat{H}_{|0} + \sum_n^{} {^S{\hat{H}}}^n_{|2} \right) \Psi
\equiv \left( \hat{H}^G + \hat{H}^M \right) \Psi = 0 \,,
\label{eq:whdewitts}
\end{equation}

\begin{equation}
\hat{H}^n_{|1} \Psi = {^S\hat{H}^n_{\_1}} \Psi = 0 \,.
\label{eq:mom1}
\end{equation}

\noindent
The indices $0$, $1$ and $2$ indicate the order with respect to the
perturbations, and $S$ the scalar part
of the total Hamiltonian (obtained by setting to 0 the vector and tensor
perturbations in \cite{hallhaw}). $\hat{H}^G$ and $\hat{H}^M$ are
respectively the gravitational and matter parts of the Hamiltonian and
$\Psi(a, \{a_n\}, \{b_n\}, \phi, \{f_n\})$ is the total matter--gravity
wave function.

We now follow a procedure analogous to the one illustrated elsewhere
\cite{broutventuri} \cite{vent90} and factorize the wave function $\Psi$ as:

\begin{eqnarray}
\Psi({a}\,, \{a_n\}\,, \{b_n\}\,, {\phi},  \{f_n\}) & = &
\prod_n \tilde{\chi}_n({a}\,, a_n\,, b_n\,, \phi, f_n)
\tilde{\psi}_n({a}\,, a_n\,, b_n) \nonumber \\
& = & \prod_n \tilde{\chi}_n \prod_l \tilde{\psi}_l \equiv
\tilde{\chi}\tilde{\psi} \,,
\label{eq:factor}
\end{eqnarray}

\noindent
which on substituting into eq. (\ref{eq:whdewitts}) leads to the
following coupled equations:

\begin{equation}
\left( \hat{H}^G + \langle \hat{H}^M \rangle \right) \tilde{\psi} =
- \langle \hat{H}_{kin}^G \rangle \tilde{\psi}\,,
\label{eq:eqpsi}
\end{equation}

\begin{equation}
(\hat{H}^M - \langle \hat{H}^M \rangle ) \tilde{\chi} + \frac{\hbar^2}{m_p^2}
(\nabla \log \tilde{\psi} )^T G \nabla \tilde{\chi} =
- ( \hat{H}_{kin}^G - \langle \hat{H}_{kin}^G \rangle ) \tilde{\chi}\,,
\label{eq:eqchi}
\end{equation}

\noindent
where  $\hat{H}_{kin}^G$ is
the gravitational kinetic energy term and the averaging procedure
$\langle \rangle$ is over all matter field configurations.
Further $G$ is a matrix depending on the diverse gravitational degrees
of freedom, $\nabla$ is a vector gradient with respect to the various
gravitational degrees
of freedom and $\nabla^T$ its transpose. It is understood that eq.
(\ref{eq:eqchi}) is evaluated where $|\tilde{\psi}|^2$ has support
\cite{vent90}.
The terms on the r.h.s. of eqs. (\ref{eq:eqpsi}) and (\ref{eq:eqchi}) are
associated with fluctuations and disappear in the BO
approximation. Analogous
equations are obtained from the constraints eq. (\ref{eq:mom1}).

We may now introduce a semiclassical approximation to the gravitational wave
function through:

\begin{equation}
\tilde{\psi}(\tilde {a}, \{a_{n}+b_{n}\})=N_G e^{\frac{i}{\hbar}S^G},
\end{equation}

\noindent
with

\begin{equation}
\tilde{a}=a e^{\frac{1}{2} \sum_n a_n^2
- 2 \sum_n \frac{n^2 - 4}{n^2 -1} b_n^2} \,,
\end{equation}

\noindent
where $S^G$ is the classical gravitational action and is solution
to the Hamilton--Jacobi equation obtained for $\hbar \rightarrow 0$
from eq. (\ref{eq:eqpsi}) in the absence of fluctuations, and the
contribution to the lower order (in $\hbar$) from $N_G$ is
negligible in our present approximation. Further, since the momentum
constraints
eq. (\ref{eq:mom1}) are associated with reparametrisation invariance and in
general reduce the number of parameters leading to a mixing of matter and
gravitation degrees of freedom \cite{wada} in contrast with our
factorization ansatz eq. (\ref{eq:factor}), we require that they be satisfied
just for the gravitational
wave function $\tilde\psi$ in the absence of matter backreaction \cite{wada}.
This then leads to the above parameter dependence in
$\tilde\psi$.

Through the above, as previously explained \cite{banks}--\cite{broutventuri},
one may introduce a conformal time $\eta$ and eq. (\ref{eq:eqchi})
becomes:

\begin{eqnarray}
(\hat{H}^M - \langle \hat{H}^M \rangle ) \tilde{\chi}
-\frac{i \hbar}{m_p^2}
(\nabla \log S^G )^T G \nabla \tilde{\chi} & = &
(\hat{H}^M - \langle \hat{H}^M \rangle ) \tilde{\chi}
-i \hbar \frac{\partial}{\partial\eta} \tilde{\chi}\nonumber \\
& = &
- ( \hat {H}_{kin}^G - \langle \hat {H}_{kin}^G \rangle ) \tilde{\chi}
\label{eq:sch1}
\end{eqnarray}

\noindent
and in particular on neglecting the fluctuation in eq. (\ref{eq:sch1})
one obtains:

\begin{equation}
\left( \hat{H}^M - i \hbar \frac{\partial}{\partial\eta} \right)
e^{-\frac{i}{\hbar} \int^\eta \langle \hat{H}^M \rangle d\eta'} \tilde{\chi}
\equiv
\left( \hat{H}^M - i \hbar \frac{\partial}{\partial\eta} \right)
\chi_s =0 \,,
\label{sch}
\end{equation}

\noindent
which is the usual evolution equation for matter.
Eq. (\ref{sch}) is then solved through an ansatz for $\chi_s$ \cite{guven}:

\begin{equation}
\chi_s = N e^{\frac{i}{\hbar}S}
\end{equation}

\noindent
where:

\begin{eqnarray}
\lefteqn{
S=S_0(a,\phi) +
} \nonumber \\ & &
\sum_n \left(
\frac{1}{2} S^{(n)}_{aa} a_n^2 +  \frac{1}{2} S^{(n)}_{bb} b_n^2 +
\frac{1}{2} S^{(n)}_{ff} f_n^2 + \right. \nonumber \\
& & \left. S^{(n)}_{ab} a_n b_n +
S^{(n)}_{af} a_n f_n + S^{(n)}_{bf} b_n f_n \right) \,,
\end{eqnarray}

\noindent
which is substituted into eq. (\ref{sch}). On equating coefficients of the
same order in $a_n \,, b_n \,, f_n$, a series of relations are then obtained
and solved for the functions $S^{(n)}(a,\phi)$.

The solutions for $\tilde\psi$ and $\tilde\chi$ are then used to obtain
expressions of physical interest. Further, since a semiclassical limit for
$\tilde\psi$ is considered, for the values of $a$, $a_n$ and $b_n$ one uses
their average (classical) values. In particular, from the matter wave function
$\chi_s$ (solution to eq. (\ref{sch})) one determines the
expectation values of the coefficients $f_n^2$ at the exit from the horizon,
obtaining with $H$ the
Hubble parameter:

\begin{equation}
\langle f_n^2 \rangle \simeq \frac{\hbar H^2}{2 n^3} \,,
\label{unpert}
\end{equation}

\noindent
which is directly related to the spectrum of density
fluctuations and corresponds to the Harrison--Zel'dovich (HZ)
spectrum (since $k=\frac{n}{2\pi a}$, with $k$ the physical wave number)
\cite{peebles}.
Further, one may estimate, using the lowest order solution of eq. (\ref{sch}),
the magnitude of the fluctuations in eq. (\ref{eq:sch1}) through:

\begin{equation}
\left( \hat{H}_{kin}^{G} - \langle \hat {H}_{kin}^G \rangle \right)
\tilde{\chi} \simeq \pm
\left[ \langle \stackrel \leftarrow {\hat H}{_{kin}^{G}} \hat {H}_{kin}^G
\rangle -
\langle \hat {H}_{kin}^G \rangle^2 \right]^{\frac{1}{2}}  \tilde{\chi} \,,
\label{fluc}
\end{equation}

\noindent
which may be interpreted as a correction to the scalar field energies and
consequently as a distortion of the unperturbed (HZ) spectrum and we shall
denote it by $\Delta \langle f_n^2 \rangle$.
One then obtains:

\begin{equation}
\frac{\Delta \langle f_n^2 \rangle}{\langle f_n^2 \rangle} = \pm
\frac{\left[ \langle \stackrel \leftarrow {\hat H}{_{kin}^{G}} \hat {H}_{kin}^G
 \rangle -
\langle \hat {H}_{kin}^G \rangle^2 \right]^{\frac{1}{2}} }{\langle \hat {H}^M
\rangle} \,,
\label{pazzia}
\end{equation}

\noindent
where the scale factor is evaluated at the exit from the horizon for each mode.

\begin{figure}
\caption{Unperturbed spectrum (HZ) and its distorsion (fluctuation
corrections $+/-$)
power spectrum at $\simeq 10^{17} GeV$ and $\simeq 10^{14} GeV$ inflation
energy.}
\label{figfig}
\end{figure}

In particular, if the spectrum is evaluated at a common time after
having re--entered the horizon, one finds that the unperturbed part behaves
as $n$, while the corrections behave as $\simeq n^{-1}$ and
$\simeq n^{2}$
for small and large $n$ respectively and are therefore most
effective both at extremely large and small scales.
Further the interval in $n$
for which our perturbative approach is valid increases as the
cosmological constant driving inflation is decreased (see Fig. \ref{figfig}).
For example, for an inflation energy $\simeq 10^{17} GeV$
and an e--folding of $100$, the HZ spectrum is
modified by our perturbation on physical scales in the interval
($1\div10^{4}$) $h^{-1} Mpc$
(we adopt the value $h = 0.5$ for the present value of
the Hubble constant $H_{0}$ in units of $100\, km \,\,s^{-1} Mpc^{-1}$).
However, from the analysis of present inflationary models \cite{TRNXX}
one obtains that standard and chaotic inflation is consistent with the
COBE results \cite{SMOO} \cite{BENN} only for energies $\simeq 10^{14} GeV$
(which is the energy scale for which the observable fluctuations exit the
horizon) and in order to obtain a higher energy one
should construct models with suitable e--folding and/or a horizon
growth during inflation steeper than that presently used \cite{TRNXX}.

With the fluctuation spectra obtained (HZ $\pm$ fluctuation corrections)
and for an inflation energy of
$\simeq 10^{14} GeV$, we have performed numerical simulations
assuming cold dark matter dominance (the so--called CDM model): the
constituents of dark matter in this model are massive particles, which
decoupled from radiation when non--relativistic or never were in thermal
equilibrium.

In the last decade the standard CDM model has shown a high predictive power
in explaining many observed properties of the large--scale galaxy
distribution. However, it is now known that this model has some serious
problems, mostly due to the high ratio of small to large--scale power. In
particular, the COBE normalization \cite{SMOO} \cite{BENN} implies excessive
velocity dispersion on $Mpc$ scales \cite{GEBE} and is unable to reproduce
the slope of the galaxy angular correlation function obtained from the APM
survey \cite{MADD}. The spectrum of the primordial fluctuations in the CDM
model is the HZ spectrum and, as mentioned, our
solutions modify this spectrum at small and very large scales.

To follow the non--linear evolution after the matter--radiation decoupling
we used a particle--mesh code \cite{MESS}.
A preliminary analysis with $N_{p} = 128^{3}$ particles and
$N_{g} = 128^{3}$ grid--points indicated that adding or subtracting
our fluctuation corrections to the HZ spectrum in the large $n$ region,
increased or decreased
respectively the mass excess or the bulk velocities with respect to
the HZ results. Therefore, we proceeded with more detailed simulations with
$N_{p} = 256^{3}$ particles on $N_{g} = 256^{3}$ grid--points, on a Cray
T3D MCA 64--8.

We ran three simulations, one with the HZ spectrum over a box of size
$128 h^{-1} Mpc$ and two with the spectrum obtained on subtracting the
fluctuation corrections from the HZ spectrum (we shall denote it by
$QF_{-}$), one over a box of size
$128 h^{-1} Mpc$ and the other over a box of size $512 h^{-1} Mpc$.

The amplitude of the primordial fluctuation spectrum is not determined
by our free theoretical parameters (that is, the inflation energy and
the e--folding number) but by the inverse of the {\it {rms}}
mass fluctuation on a sharp--edged sphere of radius $8 h^{-1} Mpc$
($\sigma_{8}$). The COBE DMR detection of large angular scale
anisotropies of the cosmic microwave background \cite{SMOO} then
fixes the normalization and makes the model completely specified.

\begin{table}
\centering
\caption{Comparison of observed and computed mass variance.}
\begin{tabular}{cccc}
$h^{-1}Mpc$ & $APM$\cite{LOVE} &  $QF_{-}$      & $HZ$ \\
$10$        & $1.05-1.47$      &  $1.10-2.08$   & $2.31-3.07$ \\
$15$        & $0.61-0.89$      &  $0.58-0.85$   & $0.99-1.19$ \\
$20$        & $0.39-0.62$      &  $0.35-0.43$   & $0.55-0.57$ \\
$25$        & $0.29-0.49$      &  $0.21-0.24$  & $0.31-0.33$ \\
$30$        & $0.18-0.33$      &  $0.13-0.16$  & $0.18-0.21$ \\
\end{tabular}
\label{sigma}
\end{table}

A detailed study of the numerical results is under way, however a
preliminary analysis of the variance of
mass reported in Table \ref{sigma} for the box of size
$128 h^{-1} Mpc$ (the range of values for $QF_{-}$ and HZ refers to
the greatest and the least of three evaluations obtained
following \cite{LOVE}) gives an idea of the relevance of
the fluctuation corrections to the development of large scale structure.
Analysis of mock galaxy catalogues obtained by such simulations should
then provide a quantitative answer concerning the possibility of solving
the two major problems of the CDM model with quantum corrections.

Further details of theoretical and numerical aspects will be presented
elsewhere \cite{BERT} \cite{Paper}.

\end{document}